\newtheorem{definition}{\textbf{Definition}}

\newtheorem{proposition}{\textbf{Proposition}}

\newcommand{\Fangxin}[1]{\textcolor{black}{#1}}

\documentclass[10pt,journal,compsoc]{IEEEtran}

\ifCLASSOPTIONcompsoc
  \usepackage[nocompress]{cite}
\else
  \usepackage{cite}
\fi
\ifCLASSINFOpdf
  \usepackage[pdftex]{graphicx}
\else
\fi
\usepackage{amsmath}
\usepackage{mathtools}
\usepackage{amssymb}
\usepackage{amsfonts}
\usepackage{soul}
\usepackage{color}
\usepackage{graphicx}
\usepackage{subcaption}
\usepackage{lipsum}
\usepackage{booktabs}
\usepackage{multirow}
\usepackage{array}
\usepackage{epstopdf}
\usepackage{cases}

\usepackage{algorithm}
\usepackage{algorithmic}
\begin{document}
%
\newcommand{\sys}{\texttt{Fumos}\xspace}
\title{LMaaS: Exploring  Pricing Strategy of  Large Model as a Service for Communication}
%
%
%

\author{Panlong~Wu,~\IEEEmembership{}
        Qi~Liu,~\IEEEmembership{}
        Yanjie~Dong,~\IEEEmembership{Member,~IEEE}
        and~Fangxin~Wang,~\IEEEmembership{Member,~IEEE}
\IEEEcompsocitemizethanks{\IEEEcompsocthanksitem Panlong Wu is with the Future Network of Intelligence Institute and the School of Science and Engineering, The Chinese University of Hong Kong, Shenzhen, Shenzhen 518172, China. E-mail: panlongwu@link.cuhk.edu.cn.
\IEEEcompsocthanksitem Qi Liu is with the School of Science and Engineering, The Chinese
University of Hong Kong, Shenzhen, Shenzhen 518172, China. E-mail: 120090027@link.cuhk.edu.cn.
\IEEEcompsocthanksitem Yanjie Dong is with Shenzhen MSU-BIT University, Shenzhen, China. E-mail: ydong@smbu.edu.cn.
\IEEEcompsocthanksitem Fangxin Wang is with the School of Science and Engineering and the
Future Network of Intelligence Institute, The Chinese University of Hong
Kong, Shenzhen and Guangdong Provincial Key Laboratory of Future
Networks of Intelligence. Email: wangfangxin@cuhk.edu.cn.}
\thanks{Manuscript received xxx; revised xxx.}}

%
%

\markboth{IEEE TRANSACTIONS ON MOBILE COMPUTING}
{Shell \MakeLowercase{\textit{et al.}}: Bare Advanced Demo of IEEEtran.cls for IEEE Computer Society Journals}
%



\IEEEtitleabstractindextext{%
\begin{abstract}
The next generation of communication is envisioned to be intelligent communication, that can replace traditional symbolic communication, where highly condensed semantic information considering both source and channel will be extracted and transmitted with high efficiency. The recent popular large models such as GPT4 and the boosting learning techniques lay a solid foundation for the intelligent communication, and prompt the practical deployment of it in the near future. Given the characteristics of ``training once and widely use'' of those multimodal large language models, we argue that a pay-as-you-go service mode will be suitable in this context, referred to as 
Large Model as a Service (LMaaS). However, the trading and pricing problem is quite complex with heterogeneous and dynamic customer environments, making the pricing optimization problem challenging in seeking on-hand solutions. 
In this paper, we aim to fill this gap and formulate the LMaaS market trading as a Stackelberg game with two steps. In the first step, we optimize the seller's pricing decision and propose an Iterative Model Pricing (\texttt{IMP}) algorithm that optimizes the prices of large models iteratively by reasoning customers' future rental decisions, which is able to achieve a near-optimal pricing solution. In the second step, we optimize customers' selection decisions by designing a robust selecting and renting (\texttt{RSR}) algorithm, which is guaranteed to be optimal with rigorous theoretical proof. Extensive experiments confirm the effectiveness and robustness of our algorithms.

\end{abstract}

}

\maketitle

\IEEEdisplaynontitleabstractindextext

%
\IEEEpeerreviewmaketitle

\ifCLASSOPTIONcompsoc
\IEEEraisesectionheading{\section{Introduction}\label{sec:introduction}}
\else
\section{Introduction}
\label{sec:introduction}
\fi

According to Shannon's theory~\cite{vaswani2017attention}, communication will experience three stages, i.e., symbolic transmission, semantic transmission and effectiveness of transmission. In the past few decades, symbolic communication has ushered an explosive development, from 1G to 5G/B5G, raising the transmission rate to even Gbps. But due to scarce spectrum resources, the room for further improvement is quite limited. 
As envisioned, the rise of semantic communication (SC)~\cite{qin2021semantic, dai2022nonlinear, dong2022semantic} sheds a light on the future of communication, pushing it to evolve to the second stage, where highly condensed semantic information will be extracted and transmitted, rather than the raw symbolic information. Semantic communication is able to empower many applications such as the industrial Internet, 3D video streaming, and the future Metaverse, with great market potential.



\begin{figure}[t]
  \centering
  \includegraphics[width=0.5\textwidth]{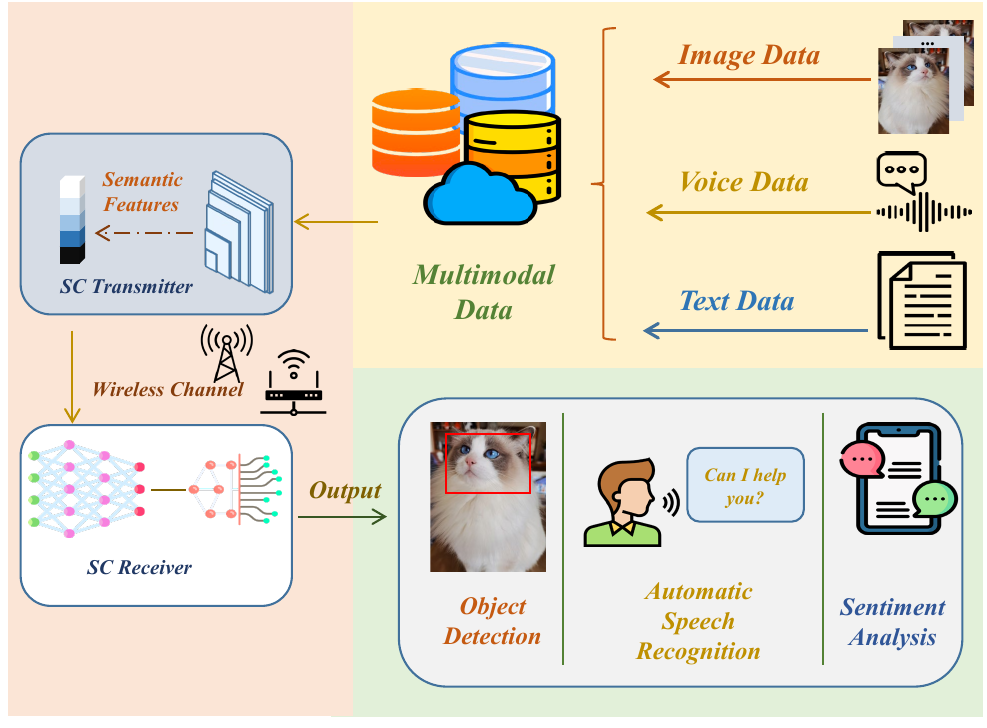}
  \vspace{-0.3cm}
  \caption{The multimodal semantic communication workflow.}\label{SC_tech}
  \vspace{-0.3cm}
\end{figure}

In semantic communication architecture, the key components are the semantic encoder and decoder~\cite{xie2021deep}, which are usually advanced neural network models, aiming to effectively extract semantic features. The encoder and decoder not only incorporate communication source and task information but also include the communication channel features, towards feature fusion to better adapt to various conditions. Fig.~\ref{SC_tech} illustrates a typical semantic communication workflow, where multimodal data will be encoded, transmitted, and decoded in a semantic manner, with the received data ready to satisfy the task requirement. The recent popular large models such as GPT models are expected to be the best choice for semantic communication codecs considering their unprecedented ability in content understanding and generalization.


Training a semantic encoder and decoder represented by large model requires huge endeavor and cost, e.g., only the training of a GPT3 model requires millions of dollars, not to mention the infrastructure cost and the labor cost. Thus, training a semantic model is unlikely to be an individual activity. On the contrary, this kind of service is very suitable to be carried out by service providers since large models are quite in line with the principle of ``training once and widely use''. In this context, the pricing strategy of the semantic model for both customers and the service provider to maximize their profit is particularly important. 

\Fangxin{In the  market, the service provider will first release some foundation models with diverse capability and resource consumption, aiming to maximize the potential revenue. Every customer is expected to choose a foundation model according to their own demand and budget to maximize his/her profit. A series of customized supervised fine-tuning, reward modeling, and reinforcement learning are then applied to the selected models to generate the final personalized models, which are then delivered to customers to support high throughput applications such as VR/AR streaming, video analytics, etc. 
}


Solving this pricing problem is difficult with unique challenges in the following two aspects: (1) \textbf{Heterogeneity in customer environment}. Different from traditional pricing problems with homogeneous clients, the pricing strategy in the semantic communication context is affected by quite heterogeneous customer environments such as transmission distance, channel condition, and energy capacity, making this problem complex. (2) \textbf{Dynamics in customer environment}. Besides heterogeneity, the customer environment is also highly dynamic with strong uncertainty, making it hard to guarantee their profit in the worst case. 
Existing works~\cite{aujla2017stackelberg, su2021game, li2019stackelberg, ding2022optimal, he2021optimal} on pricing strategy failed to fully consider the dynamics in the environments, and thus cannot guarantee the performance in the worst case. 

In this paper, we for the first time fully investigate the semantic model pricing problem between the service provider and multiple customers, and address the above challenges by modeling it as a two-step multi-follower Stackelberg game. 
In the first step, we optimize the seller's pricing decision, which is jointly determined by the renting revenue and the model training cost. We propose an Iterative Model Pricing (\texttt{IMP}) algorithm that optimizes the price of SC models iteratively by reasoning customers' future rental decisions, which is able to achieve near-optimal performance. In the second step, we optimize customers' selecting and renting decisions, which are affected by the communication cost, model rental cost, and application profit. We formulate the profit maximization problem as a two-stage robust optimization problem. We then design a novel robust selecting and renting (\texttt{RSR}) algorithm that decomposes the problem into a master problem and a subproblem, and solves them iteratively. \texttt{RSR} is able to achieve optimal selection with rigorous theoretical proof.

Extensive experiments are conducted to verify the effectiveness and robustness of our algorithms under various settings. Results show that the \texttt{RSR} algorithm outperforms other baselines by up to $43.96\%$ and \texttt{IMP} algorithm reaches a performance level of $99.61\%$ with $23.46\%$ execution time compared to near-optimal baseline.

In general, our contribution can be summarized as follows:
\begin{itemize}
     \item To our best knowledge, we are the first to propose the paradigm of Large Model as a Service (LMaaS) and comprehensively consider the pricing problem therein, identifying the unique challenges in environment heterogeneity, environment dynamics, and varying training cost. 
     \item We formulate the SC models' trading process into a Stackelberg game including a leading game and a following game. In the leading game, to tackle the seller's profit maximization problem, we propose an \texttt{IMP} algorithm to optimize the seller's model price with near-optimal performance. 
     \item In the following game, to tackle the customers' profit maximization problem, we formulate the following game into a two-stage robust optimization problem and propose a novel \texttt{RSR} algorithm to optimize customers' profit under the worst case. We prove the result will converge to global optimal solution.
     
     
\end{itemize}

\section{System Model}

\begin{table}[t]
\centering
\caption{Notations}
\label{tab:notation}
\begin{tabular}{c|l}
\hline
Notations & Descriptions \\
\hline\hline
$\tau_n$ & SC model rent duration of customer $n$ \\
$b1$ & Price of unit energy \\
$U$ & Number of SC models \\
$N$ & Number of customers \\
$A$ & Utility coefficient of customers \\
$T_{\min}$ & Minimum SC model rent duration \\
$T_{\max}$ & Maximum SC model rent duration \\
$k_{\psi_n}$ & Semantic encoding speed of SC model $\psi_n$ \\
$\beta_{\psi_n}$ & Price of SC model $\psi_n$ \\
$\hat d_n$ & Basic transmission distance of customer $n$ \\
$\hat q_n$ & Basic energy budget of customer $n$ \\
$\tilde{r}_n$ & Maximum deviation of distance \\
$\tilde{\theta}_n$ & Maximum deviation of energy budget\\
$H_{\psi_n}$ & Basic charge for SC model $\psi_n$ \\
$a_{\psi_n}$ & Semantic meaning compressibility of SC model $\psi_n$ \\
$d_n$ & Distance between the device of customer $n$ and its\\& associated base station \\
$N_n$ & Power of noise at the corresponding base station \\ & of customer $n$ \\
\hline
\end{tabular}
\end{table}

\begin{figure*}[t]
   \centering
   \includegraphics[width=0.8\textwidth]{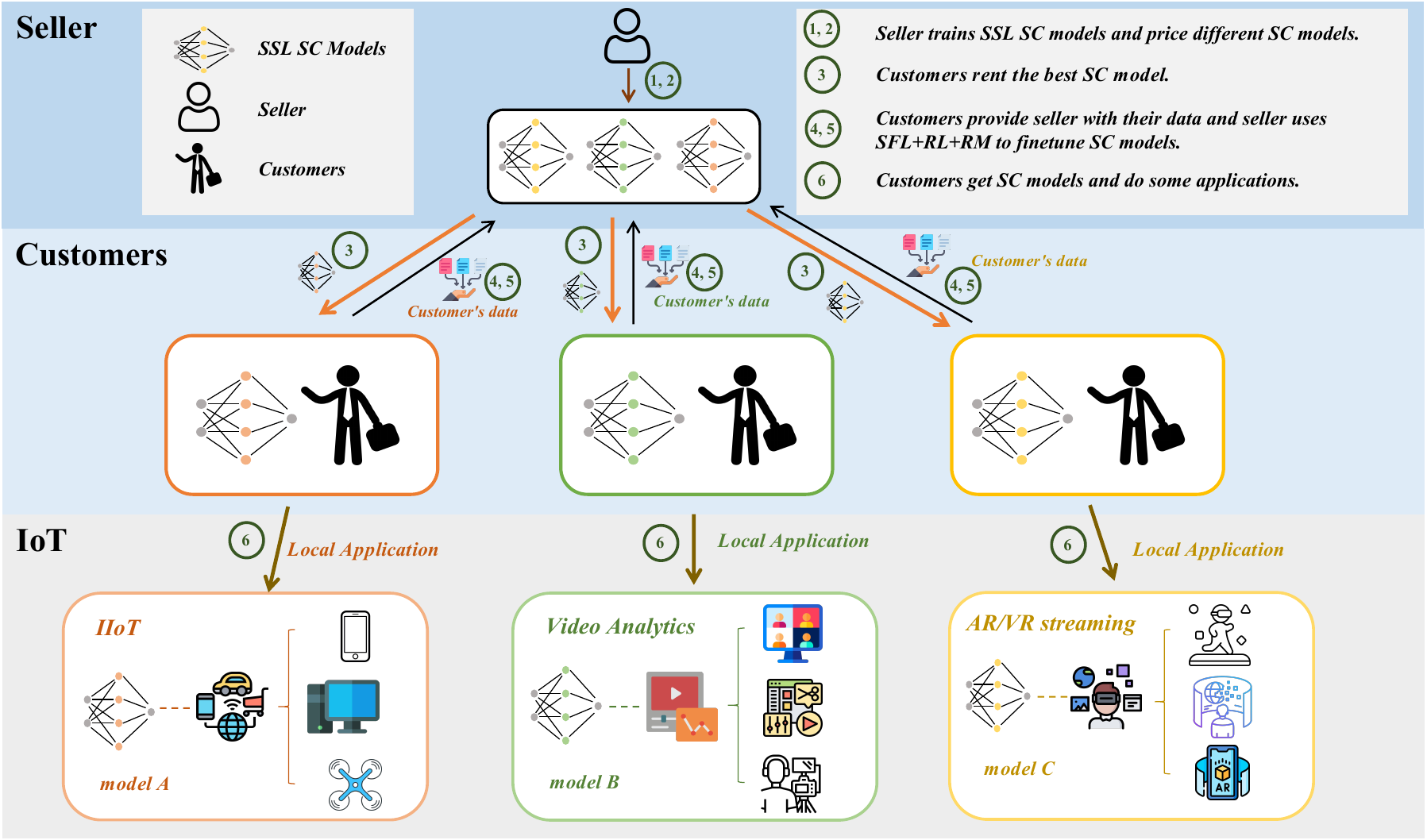}
   \caption{{The semantic communication market includes a seller and multiple customers. Seller prices different semantic communication models and customers decide model selection and rent duration.}}\label{SC_market}
   \vspace{-0.3cm}
\end{figure*}

In this section, we first articulate the SC market as a one-leader-multi-follower Stackelberg game, where both the seller and customers aim to maximize their own profits. 
As depicted in Fig. \ref{SC_market}, the SC market is composed of two main entities: the seller and customers.
The first entity is the seller that is responsible for training and providing SC models with different expenditures and capabilities. 
The second entity is a set of customers that decide the SC models selection and duration of renting the selected SC models. 
More specifically, we have $N$ customers  and $U$ SC models in the considered Stackelberg game. 
We denote the price for SC model renting by $[\beta_{u}]_{u=1}^U$, the model selection indicator by $[\psi_n]_{n=1}^N$, and the duration of customer $n$ using SC model $u$ by $[\tau_{n}]_{n=1}^{N}$.

\begin{definition}[Stackelberg Equilibrium]
The Stackelberg equilibrium (SE) denotes an action (i.e., a set of SC model prices $[\beta_{u}^*]_{u=1}^U$, SC model selection indicator $[\psi_n^*]_{n=1}^N$ and model renting duration $[\tau_{n}^*]_{n=1}^{N}$) that allows the seller and $N$ customers to maximize their own profits. 
Moreover, the entity who deviates from such action results in its own profit loss. 
The SE action satisfies
\begin{subequations}\label{eq:01}
	\begin{align}
		\Upsilon( [\beta_{u}^*, \psi_n^*, \tau_{n}^*]_{n=1, u=1}^{N, U} ) &\geq 
		\Upsilon( [\beta_{u}, \psi_n^*, \tau_{n}^*]_{n=1, u=1}^{N, U}  ) \\
		\Psi( [\beta_{u}^*, \psi_n^*, \tau_{n}^*]_{n=1, u=1}^{N, U} ) &\geq 
		\Psi( [\beta_{u}^*, \psi_n, \tau_{n}]_{n=1, u=1}^{N, U}  )
	\end{align}
\end{subequations}
where the profit of the seller is denoted by $\Upsilon$, and the profit of customers is denoted by $\Psi$. 
\end{definition}

In the SC market, the procedures of the formulated Stackelberg game are as follows: 1) the seller uses self-supervised learning (SSL) to train several SSL SC models of different architectures and scales with different  capabilities \cite{kaplan2020scaling}; 2) the seller prices different SSL SC models according to their capabilities; 3) the customers choose from the SSL SC models within their budget to maximize their own income after obtaining the renting prices of SSL SC models; 4) the customers provide the seller with their data, such as the channel state information and some historical data samples; 5) the seller utilizes the data of customers to perform supervised fine-tuning, reward modeling, and reinforcement learning to customize the SC model based on the demands of customers\footnote{For example, the data transmission for video analytics focuses more on improving the accuracy of machine learning tasks such as object detection and image classification while the data transmission for VR/AR video focuses more on improving users' quality of experience.}; 6) the customers obtain the fine-tuned SC model for subsequent tasks\footnote{For example, companies like Apple and Meta can use SC models to enhance data transmission efficiency and support AR/VR streaming which requires high throughput and low latency\cite{chen2018virtual}.}.


By considering path loss and random noise in wireless channel \cite{hao2018energy}, we can capture the energy consumption per bit due to data transmission of the semantic features by
\begin{equation}\label{eq:02}
q^{trans}_{n, u} =   d_n^{\epsilon} N_{n} a_u
\end{equation}
where $\epsilon$ is the path loss exponent; the term $d_n$ denotes the distance between the device of customer $n$ and its associated base station\footnote{In this paper, we assume that each customer owns a single device. This assumption can be easily extended to accommodate scenarios where a customer has multiple devices.}; the term $N_n$ denotes the power of additive white Gaussian noise at the base station \cite{wang2020machine}; and the term $a_u$ denotes the semantic meaning compressibility of SC model $u$. For example, a more powerful SC model can compress the original data into semantic features with smaller size.

The unit energy expenditure of semantic features transmission by customer $n$ with SC model $u$ is 
\begin{equation}\label{eq:03}
S_{n, u} = b_1 q^{trans}_{n, u}
\end{equation}
where $b_1$ is the price of unit energy. 

The total amount of original data encoded and transmitted by customer $n$ using SC model $u$ is 
\begin{align}\label{eq:04}
p_{n, u} =  k_u  \tau_n 
\end{align}
where $k_u$ represents the semantic encoding speed of SC model $u$, and 
$\tau_n$ is the renting duration of SC model by customer $n$.

Based on \eqref{eq:03} and \eqref{eq:04}, the data transmission cost is obtained as  
\begin{equation}\label{eq:05}
    C^{trans}_{n, u} = p_{n, u} S_{n, u}. 
\end{equation}

\begin{figure*}[t]
\centering
\includegraphics[width=6in, keepaspectratio]{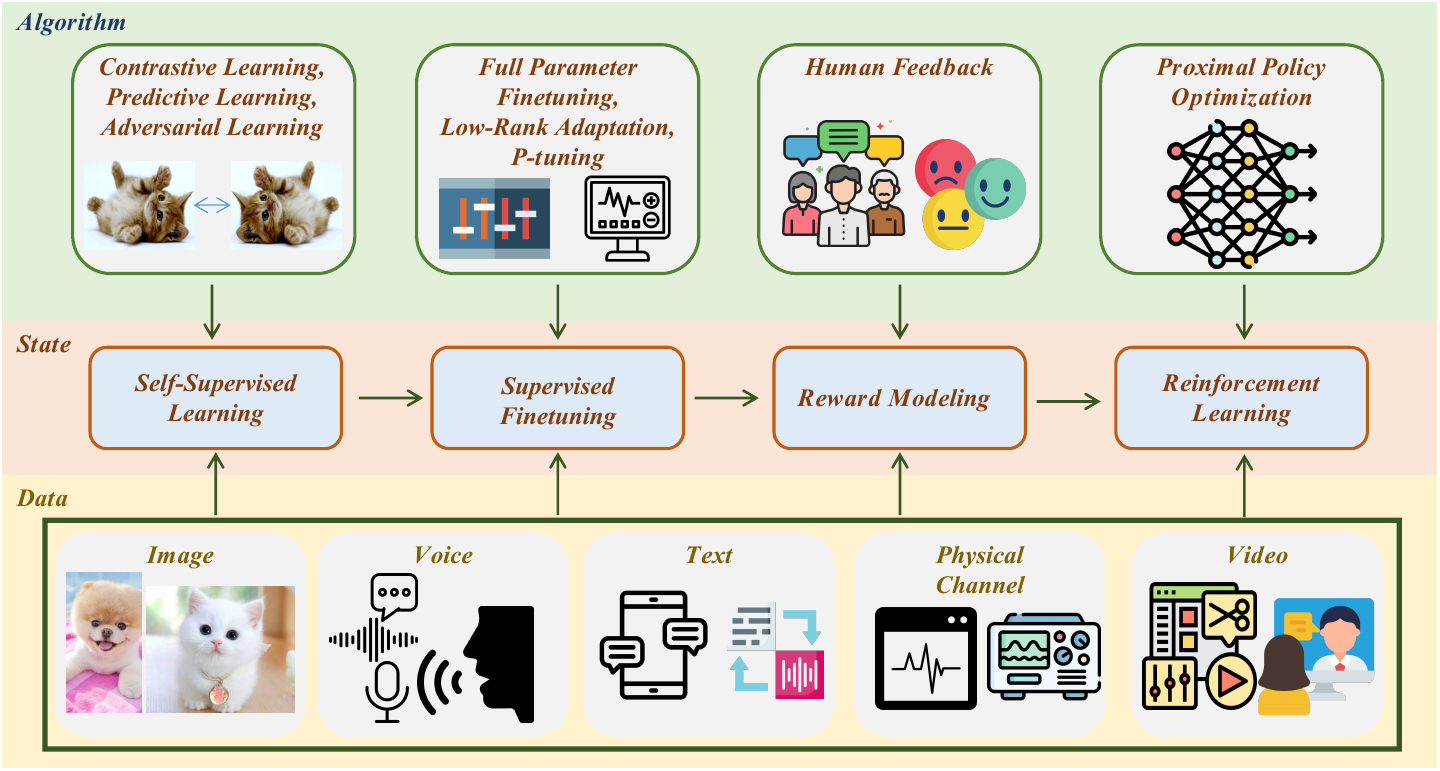}
\caption{{The training description of multimodal large language models, consisting of four parts: self-supervised learning, supervised fine-tuning, reward modeling, and reinforcement learning.}}\label{llm}
\vspace{-0.3cm}
\end{figure*}

\subsection{Utility of Customers}


The fluctuations of the deployment environment can lead to significant profit loss for customers when such fluctuations are omitted during problem formulation. Edge devices are often not fully charged when first used which will result in an uncertain energy budget\cite{dhabai2020analysis}. Moreover, the devices of customers can have uncertain distances to the corresponding base station on moving scenes such as autonomous driving and UAV monitoring.
To mitigate the profit loss, we reformulate the profit maximization problem into a two-stage robust optimization problem and propose a novel algorithm that has strong resistance to uncertainty and risk called \texttt{RSR} based on C\&CG\cite{zeng2013solving} to optimize the total profit of all customers.

More specifically, the uncertainties of transmission distance $d_n$ and energy budget $q_n$ are respectively denoted by 
\vspace{-0.1cm}
\begin{equation}\label{eq:07}
D = \{ d_n = \hat d_n + g_n\tilde{r}_n, g_n\in[0,1], \sum_{n=1}^N g_n\leq \gamma \}
\end{equation}
\vspace{-0.25cm}
and
\begin{equation}\label{eq:08}
Q = \{ q_n = \hat q_n + h_n\tilde{\theta}_n, h_n\in[0,1], \sum_{n=1}^N h_n\le \eta \}
\end{equation}
where $\hat d_n$ and $\hat q_n$ are basic transmission distance and energy budget for customer $n$; 
the term $\tilde{r}_n$ and $\tilde{\theta}_n$ are denoted as the maximum deviation of distance and energy budget; the term $\gamma$ and $\eta$ are constants that introduced to control the instability of the environment.

Due to the marginal impact of the economic market, we capture the income of each customer via a logarithmic function \cite{ding2022optimal}. Based on the \eqref{eq:02}--\eqref{eq:05}, 
We formulate a two-stage robust optimization problem as 
\begin{subequations}\label{two_stage}
\begin{align}
&\min_{[\psi_n, \tau_n]_{n=1}^N} \max_{[d_n, q_n]_{n=1}^N}  \sum_{n=1}^N [H_{\psi_n} + R_n(\psi_n, \tau_n)] \\
\mbox{s.t.}& \; p_{n, \psi_n}q_{n, \psi_n}^{trans} \le q_n, n \in [N] \\
    &\; T_{\min} \le \tau_n \le T_{\max}, n \in [N] \\
    &\; d_n\in D,q_n\in Q, n \in [N]
\end{align}    
\end{subequations}
where $\psi_n$ denotes the indicator of SC model for customer $n$; the term $A$ denotes the utility coefficient;
the term $\beta_{\psi_n}$ denotes the price of SC model $\psi_n$; the term $H_{\psi_n}$ denotes the basic charge for SC model $\psi_n$; $T_{\max}$ and $T_{\min}$ are the maximum and minimum rent duration of an SC model, $[N]$ denotes the set of indices of $N$ customers, and $R_n(\psi_n, \tau_n)$ is introduced as 
 \begin{equation}
    R_n(\psi_n, \tau_n) = C^{trans}_{n, \psi_n} +  \tau_{n}\beta_{\psi_n} - A \log(1+k_{\psi_n}\tau_n)
\end{equation}
for notation simplification.

\subsection{Utility of the seller}
The training process of a multimodal large language model is illustrated in Fig. \ref{llm}, which consists of four parts: self-supervised learning (SSL), supervised fine-tuning (SFT), reward modeling (RM), and reinforcement learning (RL). The functionalities of the four parts are described as 
\begin{itemize}
\item The SSL phase incorporates techniques like contrast learning and predictive modeling. 
While the supervised learning relies on labeled data that demand for significant expenditure of human and financial resources, the SSL has the advantage of leveraging vast amounts of unlabeled data. 
For instance, GPT3 \cite{brown2020language} is trained over 450 billion byte-pair-encoded tokens gathered from various sources, including Common Crawl and Wikipedia. 
\item In the SFT phase, the model is further trained on specific tasks or domains. SFT involves training the model with labels on a more specific dataset that is carefully generated or curated to learn task-specific behaviors. In this part, various algorithms can be used such as Low-Rank Adaptation (LoRA) \cite{hulora}, P-tuning \cite{liu2022p} to do fine-tuning with low computation cost.
\item RM phase can be employed to further enhance the behavior of the model. Instead of solely relying on supervised labels, reward signals are utilized to provide feedback on the quality of the generated outputs. These reward signals can stem from various sources, including human feedback such as user ratings or other objective metrics.
\item In the RL phase, techniques such as Proximal Policy Optimization (PPO)\cite{schulman2017proximal} can be utilized to optimize the performance of the model. The model generates responses, and it adapts its parameters based on the reward signal to enhance the quality of its outputs.
\end{itemize}


During training a SC model, the SFT, RM, and RL phases are based on the output of the SSL phase.
Moreover, different customers require different SFT, RM, and RL phases based on the same SSL phase. 
Thus, we define the cost per use of the SC model $u$ at the seller as 
\begin{equation}\label{eq:10}
cost^s_{u} = b_{2,u} N_u([\beta_{u}]_{u=1}^U) + b_3
\end{equation}
where the term $b_{2,u}$ is the cost of SC model $u$ in SFT, RM, and RL phases; the term $b_3$ is the cost of SSL phase; and $N_u([\beta_{u}]_{u=1}^U)$ denotes the number of customers renting SC model $u$.
More specifically, the value of $N_u([\beta_{u}]_{u=1}^U)$ is obtained as 
\begin{align}\label{eq:11}
\mathbf{1}_{n, u} = \left\{
\begin{array}{ll}
    1, &\psi_n^* = u \\
    0, &\mbox{otherwise.}
\end{array}\right.
\end{align}

Denote the optimal solution to \eqref{two_stage} by $[\psi_n^*, \tau_n^*]_{n=1}^N$. 
We observe from \eqref{two_stage} that the optimal renting duration $[\tau_n^*]_{n=1}^N$ is a function of renting prices $[\beta_u]_{u=1}^U$ of SC models. 
After observing the renting duration $[\tau_n^*]_{n=1}^N$ and SC model selection indicator $[\psi_n^*]_{n=1}^N$, the objective of the seller is to maximize its own profit by deciding the renting prices of SC models as
\begin{equation}\label{eq:11}
\max_{[\beta_{u}]_{u=1}^U} \sum_{n=1}^N \tau_n^*([\beta_{u}]_{u=1}^U) \beta_{\psi_n^*} - \sum_{u=1}^U cost^s_{u}
\end{equation}
where the term $\tau_n^*([\beta_{u}]_{u=1}^U) \beta_{\psi_n^*}$ is the collected revenue from customer $n$.

\section{Algorithm Design}

\subsection{Customer Algorithm Design}
The proposed two-stage robust optimization problem is decomposed into a master problem ($\cal MP$) and a subproblem ($\cal SP$). 
As shown in Algorithm \ref{RSR}, the \texttt{RSR} procedures obtain the optimal solution by iteratively solving the $\cal MP$ and the $\cal SP$.
The $\cal MP$ is
\begin{subequations}
\begin{align}
 &\min_{[\psi_n]_{n=1}^N, \alpha} \sum_{n=1}^N H_{\psi_n} + \alpha\\
\!\! \mbox{s.t.}& \sum_{n=1}^N[ C^{trans, l}_{n, \psi_n} \!+\! \tau_{n}^l \beta_{\psi_n} \!\!-\! A\log(1 \!+\! k_{\psi_n}\!\tau_{n}^l)] \!\le\! \alpha, l \!\in\!  \mathbf{O}\!\!\!\\
\!\! & k_{\psi_n}\tau_n^l d_{n,l}^{*\epsilon} N_n a_{\psi_n} - q_{n, l}^*\leq 0, n \in [N],  l \leq K \\
&T_{\min}  \le \tau_n^l \le T_{\max}, n \in [N], l \leq K\\
&C^{trans, l}_{n, \psi_n} =  k_{\psi_n} \tau_n^l b_1 d_{n,l}^{*\epsilon} N_n a_{\psi_n}, n \in [N], l \leq K
\end{align}
\end{subequations}
where the term $\alpha$ denotes the upper bound of the objective functions of all the $K$ iterations; 
the term $l$ denotes the iteration index; 
the term $\mathbf{O}$ denotes the set that contains all the index of the iteration that the $\cal SP$ is feasible and the term $K$ denotes the total iteration number; 
and, the term $d_{n,l}^*$ and $ q_{n^,l}^*$ denote the optimal solution get from the $\cal SP$ in the $l$-th iteration. 

\begin{algorithm}[!htb]\small
    \caption{Robust Selecting and Renting (\texttt{RSR})  Algorithm}\label{RSR}
    \label{alg:AOS}
    \renewcommand{\algorithmicrequire}{\textbf{Initiation:}}
    \begin{algorithmic}[1]
        \REQUIRE Set error tolerance $\epsilon$, set $K=0$, set $UB=+\infty$, $LB = -\infty$, $\mathbf{O}=\emptyset$. 
        \WHILE{$UB-LB>\frac{\epsilon}{|LB|} $}
        \vspace{0.1cm}
            \STATE Solve the ${\cal MP}$ to obtain the optimal solution $([\beta_{u,K+1}^*]_{u=1}^U, \alpha_{K+1}^*, [\tau_n^{1*}]_{n=1}^{N}, \ldots, [\tau_n^{K*}]_{n=1}^{N})$.
            \STATE Update $LB = \alpha_{K+1}^* + \sum_{n=1}^N H_{\psi_n^*, K+1}$.
            \STATE Obtain the dual problem for the inner problem of $\cal SP$.
            \STATE Update $UB \!=\! \min\{UB,\!\mathcal{Q}([\psi_{n,K+1}^*]_{n=1}^N) \!+\! \sum_{n=1}^N \!\!H_{\psi_n^*, K+1} \}$ by solving $\cal SP$.
            \IF {$\mathcal{Q}([\psi_{n,K+1}^*]_{n=1}^N) < +\infty$}         
                \STATE create new variables $([\tau_n^{K+1}]_{n=1}^{N})$ and add the following constraints to  ${\cal MP}$
                \begin{equation*}
                \begin{split}
                & \sum_{n=1}^N[ C^{trans, K+1}_{n, \psi_n} \!+\! \tau_{n}^{K+1} \beta_{\psi_n} \!\!-\! A\log(1 \!+\! k_{\psi_n}\!\tau_{n}^{K+1})] \!\le\! \alpha \\
                & k_{\psi_n}\tau_n^{K+1} d_{n,K+1}^{*\epsilon} N_n a_{\psi_n} - q_{n,K+1}^*\leq 0, n \in [N] \\
                & T_{\min}  \leq \tau_n^{K+1} \le T_{\max}, n \in [N]
                \end{split}
                \end{equation*}
                \STATE Set $K \gets K+1$, $\mathbf{O} = \mathbf{O}\cup \{K+1\}$.
            \ENDIF
            \IF {$\mathcal{Q}([\psi_{n,K+1}^*]_{n=1}^N) = +\infty$}
                \STATE add the following constraints to $\mathbf{\mathcal{MP}}$
                \begin{equation*}
                \begin{split}
                & k_{\psi_n}\tau_n^{K+1} d_{n,K+1}^{*\epsilon} N_n a_{\psi_n} - q_{n,K+1}^*\leq 0, n \in [N] \\
                & T_{\min}  \leq \tau_n^{K+1} \le T_{\max}, n \in [N]
                \end{split}
                \end{equation*}
                \STATE Set $K$ $\gets$ $K+1$.
            \ENDIF
        \ENDWHILE
    \end{algorithmic}
\end{algorithm}

Given the SC model decision from the $\cal MP$, the $\cal SP$ is to identify the worst case. 
The $\cal SP$ is set to positive infinite when infeasible. 
The $\cal SP$ can be expressed as
\begin{subequations}
\begin{align}
{\cal Q}(\psi_n^*) = 
&\max_{[d_n, q_n]_{n=1}^N} \min_{[\tau_n]_{n=1}^N} \sum_{n=1}^N R_n(\psi_n^*, \tau_n) \\
\mbox{s.t.}&\;  k_{\psi_n^*}\tau_n d_n^\epsilon N_n a_{\psi_n^*} - q_n \leq 0,  n \in [N]\\
&\; T_{\min} \le \tau_n \leq T_{\max},  n \in [N].
\end{align}    
\end{subequations}

The subproblem is a bi-level optimization problem and we transform the inner problem of $\cal SP$ into its dual problem and absorb it into the outer problem to form an equivalent single level optimization problem.

\begin{proposition}
For the inner problem of the $\cal SP$, the dual problem and the primal problem have the same optimal value. 
\end{proposition}
\vspace{-0.05cm}

\begin{IEEEproof}
The proposition is proven by demonstrating its convexity and satisfying the Slater's condition.
\begin{itemize}
\item Function $-\sum_{n=1}^N A\log(1+k_{\psi_n^*}\tau_n)$ and $\sum_{n=1}^N(C^{trans}_{n, \psi_n^*} + \tau_n \beta_{\psi_n^*})$ are convex functions w.r.t $\tau_n$; 
\item All the constraints are linear constraints;
\item According to the composition rule, the objective function in the inner problem of the subproblem is a convex 

\vspace{-0.1cm}
function;

\item Define $\mathcal{D}\!=\!\!\{\tau_n | T_{\min}\!\! \leq \!\!\tau_n \!\!\!\leq 
\vspace{0.1cm}
\!\!\!T_{\max}\mathop{\cup}\tau_n \leq \frac{q_n}{k_{\psi_n^*}d_n^\epsilon N_n a_{\psi_n^*}} \}$.
There exists a feasible $\tau_n$ satisfying $\tau_n \in \mathbf{relint\mathcal{D}}$ 
with  $k_{\psi_n}^*\tau_n d_n^\epsilon N_n a_{\psi_n^*}  -  q_n < 0, T_{\min} < \tau_n < T_{\max}$, $ n \in [N]$. 
Therefore, the Slater's condition is satisfied\cite{boyd2004convex}; 
\item Strong duality holds for the inner problem of the subproblem, resulting in the same optimal value as its dual problem.
\end{itemize}
We complete the proof. 
\end{IEEEproof}

The Lagrangian of the inner problem of the $\cal SP$ is 
\begin{equation}
\begin{split}
& L([\tau_n,\lambda_n,\mu_n,\theta_n]_{n=1}^N) \\
& = \sum_{n=1}^N R_n({\psi_n^*, \tau_n}) +\sum_{n=1}^N\lambda_n(k_{\psi_n^*}\tau_n d_n^\epsilon N_n a_{\psi_n^*}  -  q_n)\\
&\hspace{0.5 cm} +\sum_{n=1}^N\mu_n(\tau_n-T_{\max})+\sum_{n=1}^N\theta_n(T_{\min} - \tau_n).
\end{split}
\end{equation}

Construct Lagrange dual function
\begin{equation}
    g([\lambda_n,\mu_n,\theta_n]_{n=1}^N) = \inf_{[\tau_n]_{n=1}^N} L([\tau_n,\lambda_n,\mu_n,\theta_n]_{n=1}^N).
\end{equation}

Based on the definition of $R_n(\psi_n, \tau_n)$, we obtain the expression of $\tau_n$ as 
\begin{equation}\label{eq:17}
\tau_n = \frac{-A}{\theta_n-\mu_n - k_{\psi_n^*}d_n^\epsilon N_n a_{\psi_n^*}(b_1+\lambda_n)-\beta_{\psi_n^*}} - \frac{1}{k_{\psi_n^*}}.
\end{equation}

According to the Karush-Khun-Tucker condition, we can transform the $\cal SP$ into the following single level optimization problem.

\begin{subequations}
\begin{align}
&\max_{[d_n, q_n]_{n=1}^N} \sum_{n=1}^N R_{n}( \psi_n^*, \tau_n )\\
\mbox{s.t.}&\; k_{\psi_n^*}\tau_n d_n^\epsilon N_n a_{\psi_n^*} - q_n \leq 0,  n \in [N]\\
&\; T_{\min}  \leq \tau_n \le T_{\max},  n \in [N]\\
&\; (k_{\psi_n^*}\tau_n d_n^\epsilon N_n a_{\psi_n^*} - q_n)\lambda_n = 0,  n \in [N]\\
&\; (\tau_n-T_{\max})\mu_n = 0,  n \in [N]\\ 
&\; (T_{\min} - \tau_n)\theta_n = 0,  n \in [N] \\
&\; \lambda_n, \mu_n,\theta_n \ge 0,  n \in [N]\\
&\eqref{eq:07}, \eqref{eq:08}, \mbox{ and } \eqref{eq:17}
\end{align}
\end{subequations}



\begin{figure*}[h]
\centering
\begin{minipage}{0.32\textwidth}
\centering
\includegraphics[width=\textwidth]{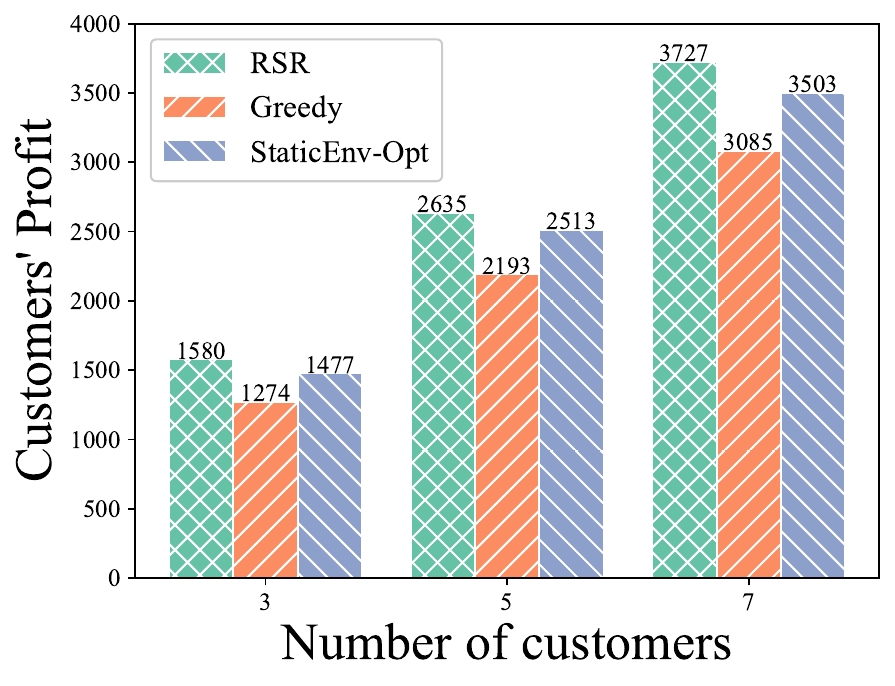}
\caption{Profit of customers with different \# of customers}
\label{fig:sub1}
\end{minipage}%
\hspace{0.1 cm}
\begin{minipage}{0.32\textwidth}
\centering
\includegraphics[width=\textwidth]{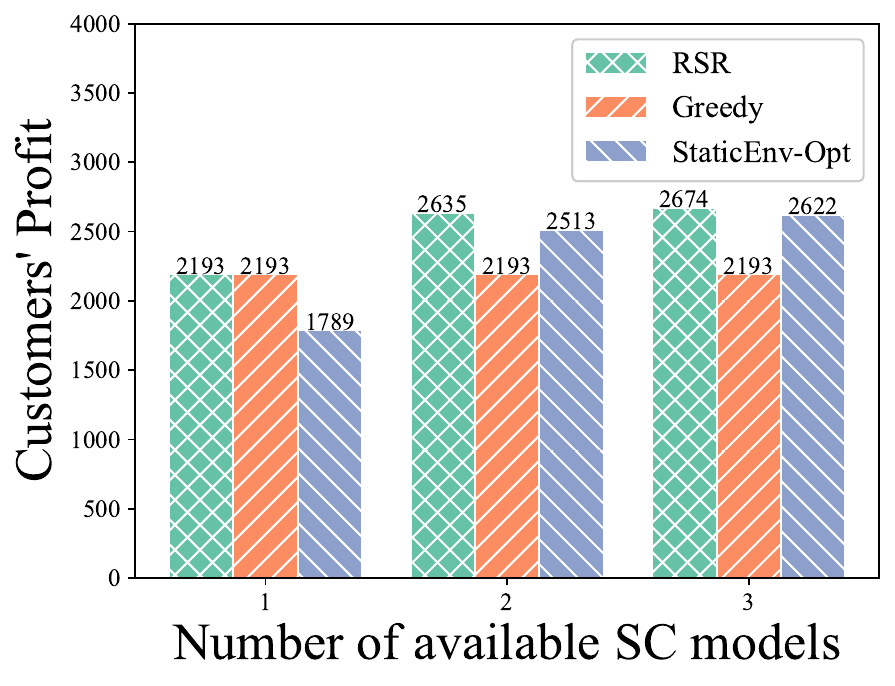}
\caption{Profit of customers with different \#  of  SC models}
\label{fig:sub2}
\end{minipage}%
\hspace{0.1 cm}
\begin{minipage}{0.32\textwidth}
\includegraphics[width=\textwidth]{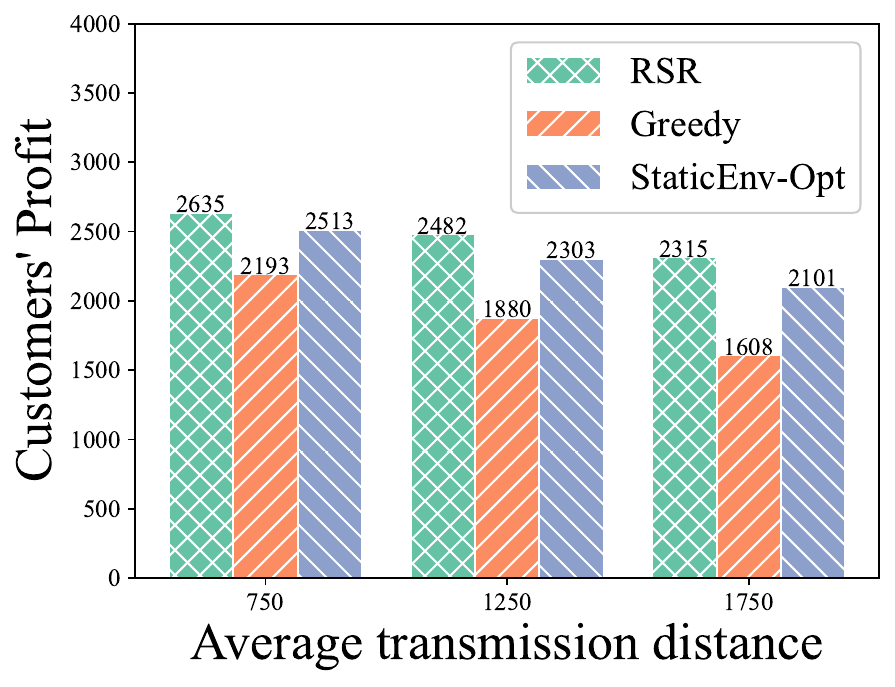}
\caption{Profit of customers with different transmission distance $d_n$}
\label{fig:sub3}
\end{minipage}
\vspace{-0.3cm}
\end{figure*}

\begin{proposition}
The proposed \texttt{RSR} algorithm guarantees the existence of a small $\epsilon > 0$ such that  $UB - LB < \frac{\epsilon}{|LB|}$. This implies that \texttt{RSR} algorithm is able to provide a tight approximation of the optimal solution within a small margin of error $\frac{\epsilon}{|LB|}$.
\end{proposition}

\begin{IEEEproof}
 In \texttt{RSR} algorithm, $LB =  \alpha_{K+1}^* + \sum_{n=1}^N H_{\psi_n^*, K+1}$ can be considered as a relaxation of the original optimization problem because of only considering part of the scenes. This lower bound will rise with more scenes incorporated in the subproblem. The procedure of getting $UB = \min\{UB,\mathcal{Q}([\psi_{n,K+1}^*]_{n=1}^N) + \sum_{n=1}^N H_{\psi_n^*, K+1} \}$ is equivalent to splitting the original problem into two independent problems and solving them separately and then take the min of the existing value and this value. In each iteration of \texttt{RSR} algorithm, the upper bound is non-increasing. With the increase of iteration number, $UB - LB < \frac{\epsilon}{|LB|}$ will be reached.
\end{IEEEproof}

\begin{algorithm}[t]
    \caption{Iterative Model  Pricing (\texttt{IMP}) Algorithm}\label{alg:REND}
    \label{alg:AOS}
    \renewcommand{\algorithmicrequire}{\textbf{Input:}}
    \renewcommand{\algorithmicensure}{\textbf{Output:}}
    \begin{algorithmic}[1]
        \REQUIRE $max\_num$  
        \ENSURE best price $[\beta_u^{*}]_{u=1}^U$    
        
        \STATE  Initialize feasible prices for all models $[\beta_u^{(0)}]_{u=1}^U$. Get the original SC model selection and rent duration of customers $[\psi_n^{(0)}]_{n=1}^N, [\tau_{n}^{(0)}]_{n=1}^{N}$ and calculate the seller's initial profit $\Upsilon( [\beta_{u}^{(0)}, \psi_n^{(0)}, \tau_{n}^{(0)}]_{n=1, u=1}^{N,U} )$.
        
        \WHILE{$\upsilon < max\_num$}
            \STATE Add prices of SC models by adding an $\delta$ to $\zeta^{(\upsilon)}$ and broadcast them to customers, then customers decide best SC models and rent duration given prices of SC models. Then the seller calculates its profit.
            \STATE Reduce prices of SC models by reducing an $\delta$ to  $\zeta^{(\upsilon)}$ and broadcast them to customers, then customers decide best SC models and rent duration given prices of SC models. Then the seller calculates its profit.
            
            \STATE Update the price of SC models $[\beta_u^{(\upsilon)}]_{u=1}^U$  through equation (\ref{app_1}) and (\ref{app_2}).
            \STATE Set $\upsilon \gets \upsilon+1$.
        \ENDWHILE
    
    \end{algorithmic}
\end{algorithm}

\subsection{Seller Algorithm Design}
The goal of the seller is to maximize its profit by adjusting the unit price per time of SC models that they lend to customers. As the leader in the Stackelberg game,  the seller has the first-mover advantage and can set the SC model price first. Customers, upon receiving the prices of different SC models, then decide the best SC models and rent duration to maximize their profit.

To address the seller pricing challenge, we propose an \texttt{IMP} algorithm to iteratively improve the seller's strategy. We set $\beta_{\psi_n}=\zeta\psi_n+\eta$ and adjust SC model price by changing $\zeta$. By adjusting the lending prices and using the profit as feedback, the seller can infer the direction of improvement of its profit and refine SC model prices.
The pseudocode for \texttt{IMP} algorithm is presented in Algorithm \ref{alg:REND}. In each iteration, the seller increases the price of SC models by a small value  and broadcasts to customers. Next, \texttt{RSR} algorithm is applied to obtain the best SC models selection and rent duration decision of customers given that prices. With this response, the seller can calculate its profit. Then seller decreases the prices of SC models by a small value and broadcasts to customers and calculates its profit. Finally, seller updates SC models prices through equation (\ref{app_1}) and (\ref{app_2})

\begin{figure*}[htbp]
  \centering
  \begin{minipage}[t]{0.32\textwidth}
    \centering
    \includegraphics[width=\textwidth]{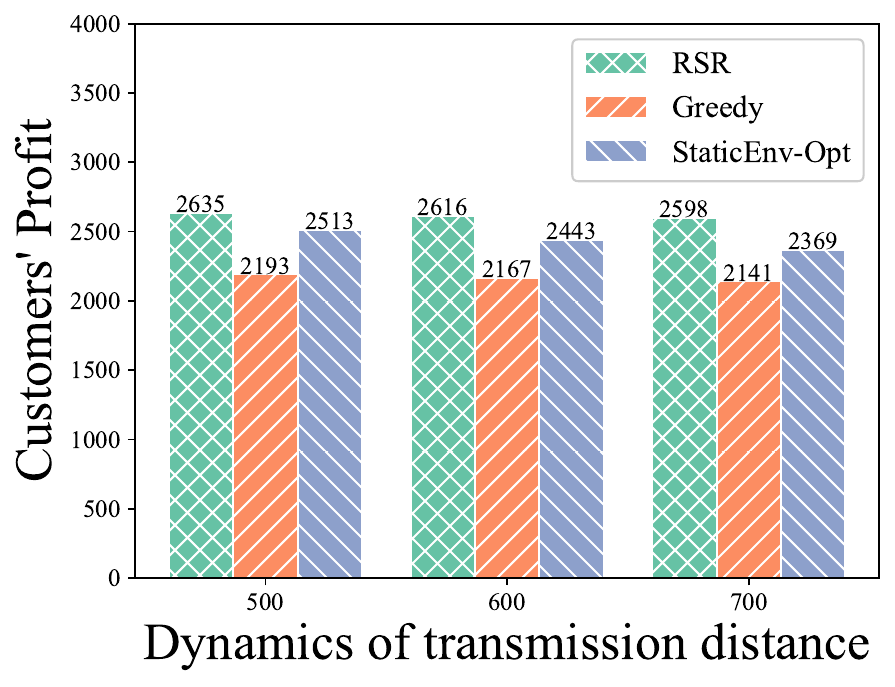}
    \caption{Profit of customers with different maximum transmission distance fluctuation $\tilde{r}_n$}
    \label{fig:cust_diff_disturbance}
  \end{minipage}
  \hfill
  \begin{minipage}[t]{0.32\textwidth}
    \centering
    \includegraphics[width=\textwidth]{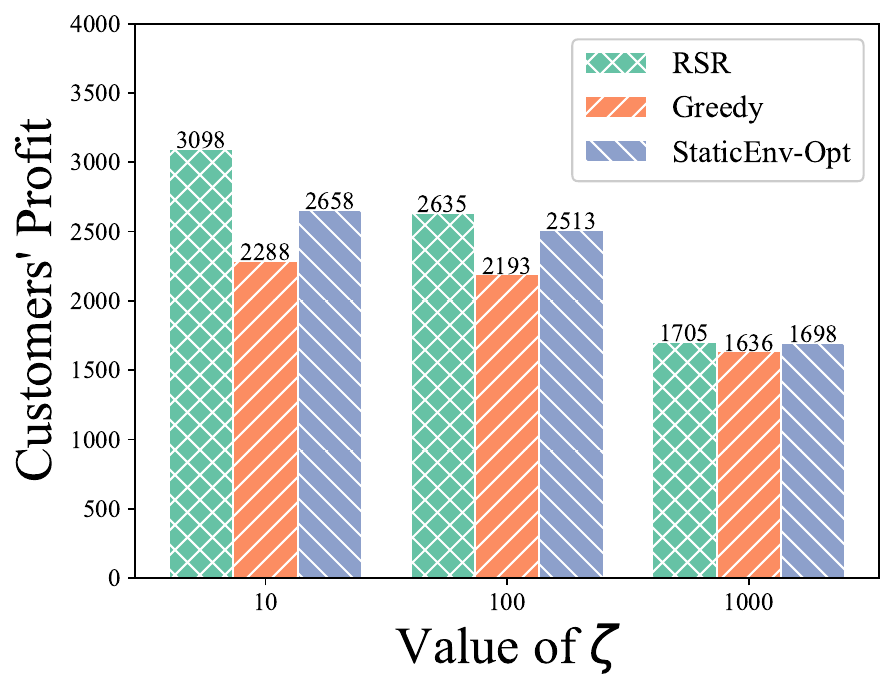}
    \caption{Profit of customers with different SC models prices}
    \label{fig:figure2}
  \end{minipage}
  \hfill
  \begin{minipage}[t]{0.32\textwidth}
    \centering
    \includegraphics[width=\textwidth]{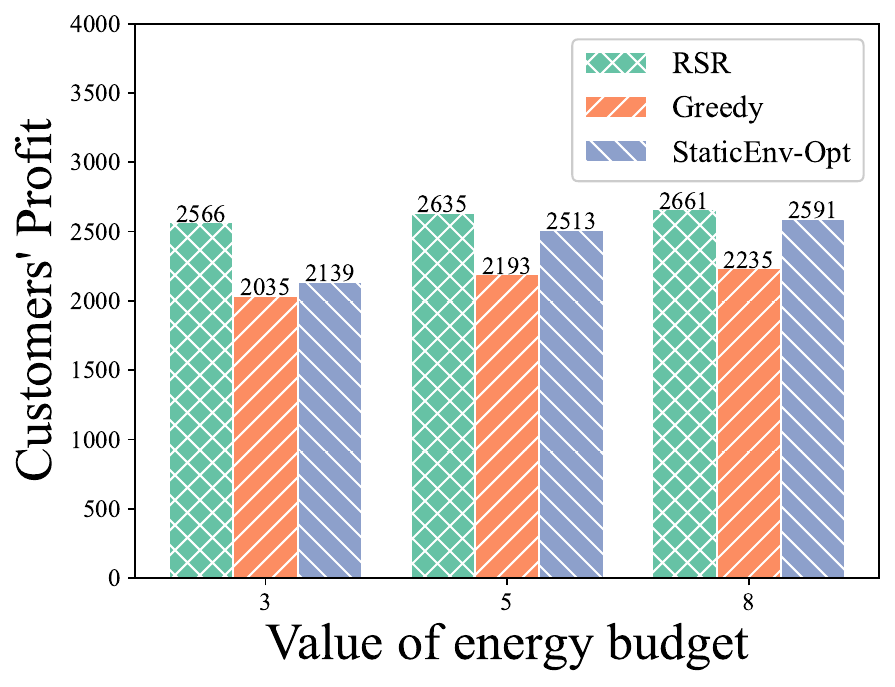}
    \caption{Profit of customers with different energy budget $q_n$}
    \label{fig:cust_energy}
  \end{minipage}
  \vspace{-0.3cm}
\end{figure*}

\begin{equation}\label{app_1}
    \zeta^{(\upsilon+1)} = \zeta^{(\upsilon)} + \chi (\frac{\partial \Upsilon}{\partial \zeta^{(\upsilon)}})
\end{equation}
where $\chi$ is the step size to control the speed of price changing; the term $\upsilon$ denotes the iteration number. In practice, we use the numerical result to approximate the partial derivative.

\begin{equation}\label{app_2}
    \frac{\partial \Upsilon}{\partial \zeta^{(\upsilon)}} \approx 
    \frac{\Upsilon(\zeta^{(\upsilon)}+\delta)-\Upsilon(\zeta^{(\upsilon)}-\delta)}{2\delta} 
\end{equation}
where $\delta$ is a small constant. We set multiple start prices to better reach near-optimal prices.
\texttt{IMP} algorithm terminates when the number of iterations reaches the maximum iteration number. 

\section{Performance Evaluation}
In this section, we conduct comprehensive experiments to verify the effectiveness of our proposed algorithm under different settings. We also discuss how different configurations can affect the results of our experiments.

\subsection{Experiments Setup}
The noise power is -77 dBm\cite{wang2020learning}. The value of path loss exponent $\epsilon$ is 3. 
The value of the utility coefficient $A$ is 100. 
The value of the price of unit energy $b_1$ is 100.
\footnote{The parameter seting is just one example, the efficiency and superior performance of our algorithm holds for all settings.}
The maximum deviation of the energy budget is set to 2W. The maximum deviation of the semantic features transmission distance is set to 500m. The minimum rent duration and maximum rent duration are set to 0 and 100 units of time. The default average semantic features transmission distance $d$ is set to 750m. The default customer number is set to 5 in the customers’ profit evaluation part and 3 in the seller’s profit evaluation part. The default SC model number is set to 2. The default energy budget is set to 5W. We assume that SC models with larger model indexes have more powerful capabilities. Semantic encoding speed $k_u$ decreases linearly with the model index $u$ and semantic meaning compressibility $a_u$ is inversely proportional to model index $u$.

\subsection{Customers' Profit Evaluation.}
To verify the effectiveness of the proposed \texttt{RSR} algorithm, we compare customers' total profit with the following baselines.
\begin{itemize}
    \item \textit{Greedy}: Customers rent the cheapest SC model and optimize the rent duration according to the ability and consumption of the SC model.
    \item \textit{Static Environment Optimal Algorithm (StaticEnv-Opt)}\cite{ding2022optimal}: Customers select the SC model and decide the rent duration according to the deterministic case by ignoring the fluctuation of the environment. 
\end{itemize} 

\textbf{Impact of the number of customers}.
Fig. \ref{fig:sub1} illustrates customers' total profit under different number of customers. Customer numbers are set to 3, 5, and 7 respectively. Results show that \texttt{RSR} algorithm achieves the highest profit in all situations. It has a higher profit from $20.15\%$ to $26.7\%$ compared to Greedy algorithm in all three settings and it has a higher profit from $4.85\%$ to $6.97\%$ compared to StaticEnv-Opt algorithm in all three settings. 

\begin{figure*}[htbp]
\centering
\begin{minipage}{0.32\textwidth}
 \centering
        \includegraphics[width=\textwidth]{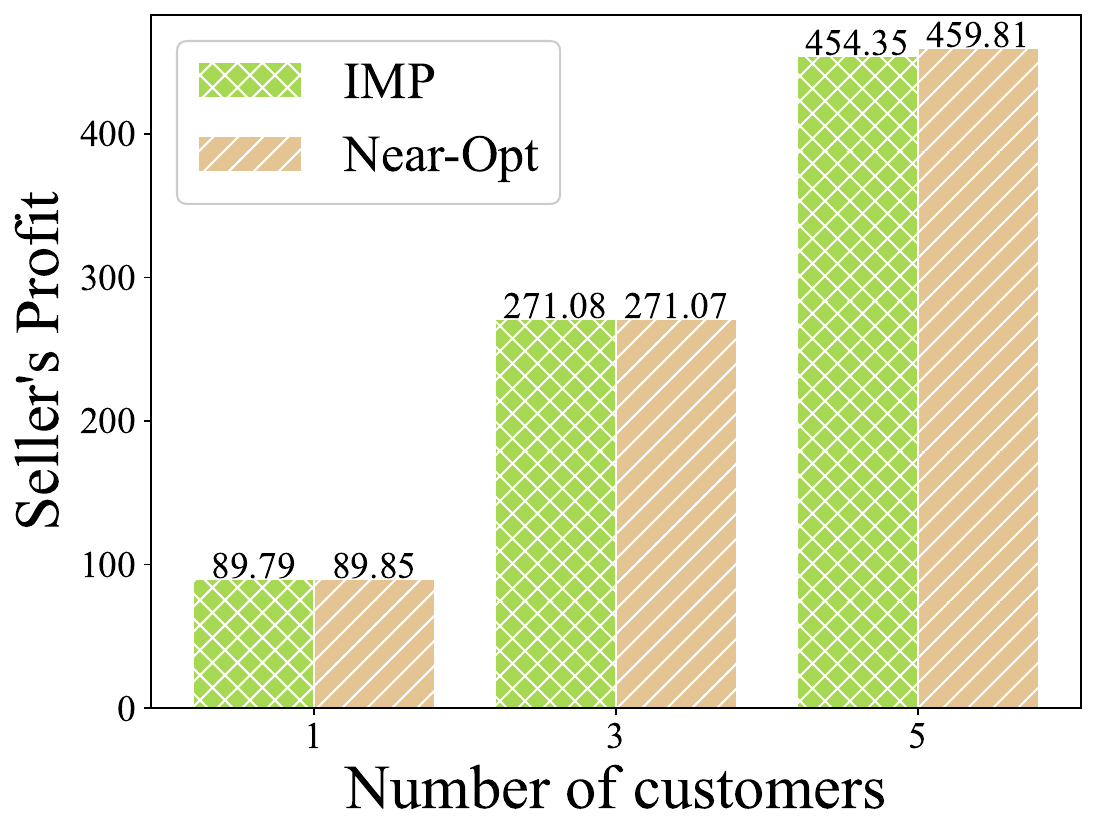}
        \caption{Profit of seller with different \# of customers}
        \label{fig:sub4_1}
\end{minipage}%
\hspace{0.1 cm}
\begin{minipage}{0.32\textwidth}
    \centering
        \includegraphics[width=\textwidth]{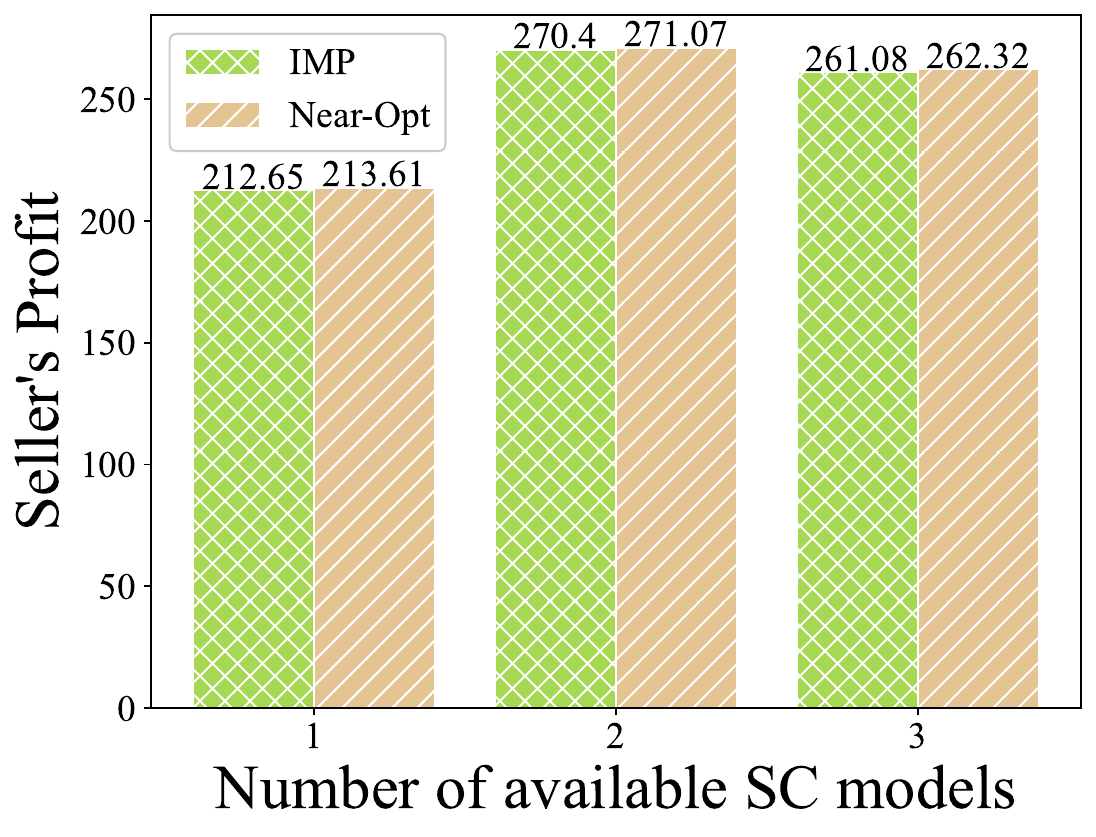}
        \caption{Profit of seller with different \# of SC models}
        \label{fig:sub4_2}
\end{minipage}%
\hspace{0.1 cm}
\begin{minipage}{0.32\textwidth}
    \centering
        \includegraphics[width=\textwidth]{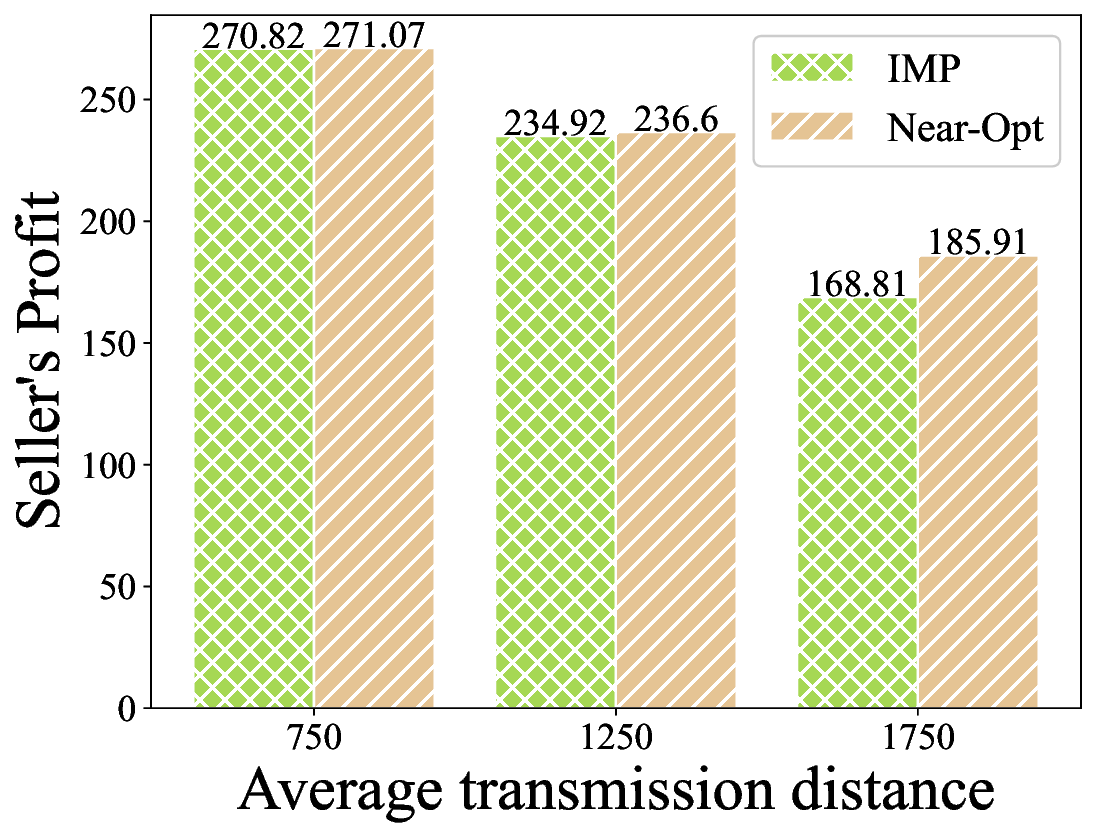}
        \caption{Profit of seller with different transmission distance $d_n$}
        \label{fig:sub4_3}
\end{minipage}
\vspace{-0.3cm}
\end{figure*}

\textbf{Impact of the number of available SC models}.
Fig. \ref{fig:sub2} indicates customers' total profit under different number of available SC models. Available SC models are set to be 1, 2, and 3 respectively. Customers have a maximum profit gain of $21.93\%$ and $22.58\%$ compared to Greedy and StaticEnv-Opt algorithms.
When the available SC model number is one, the profit of \texttt{RSR} and Greedy algorithm are the same because of choosing the same SC model and doing optimal rent duration decision. The profits of \texttt{RSR} and StaticEnv-Opt algorithm show a non-decrease trend with the increase of available SC models while the results of Greedy algorithms stay the same. This is because customers can choose more cost-effective SC models in \texttt{RSR} and StaticEnv-Opt algorithms according to their own budgets when there are multiple SC models in the market.

\textbf{Impact of semantic features transmission distances}.
We set the semantic features transmission distance to be 750, 1250, and 1750 meters to test the effectiveness of our algorithm in different transmission distances.
Fig. \ref{fig:sub3} shows customers' total profit decreases with the increase of average semantic features transmission distances. This is due to longer distances will cause larger semantic features transmission energy costs thus will harm customers' profit. Our algorithm has a profit loss gain of $20.15\%, 32.02\%$, and $43.96\%$ compared to Greedy algorithm when the transmission distance is 750, 1250, and 1750 meters. Compared to StaticEnv-Opt algorithm, our algorithm has a profit gain of $4.85\%, 7.77\%$, and $10.18\%$ under all three transmission distance settings.

\textbf{Impact of dynamics of semantic features transmission distances}.
We illustrate the profit of customers at different levels of dynamics in semantic features transmission distances. in Fig. \ref{fig:cust_diff_disturbance}. Results show that \texttt{RSR} algorithm achieves a higher profit from $20.15\%$ to $21.34\%$ compared to Greedy algorithm and achieves a higher profit from $4.85\%$ to $9.66\%$ compared to StaticEnv-Opt algorithm. It can be observed that with the increase of dynamics of the distance, \texttt{RSR} algorithm has more advantages than the two baseline methods. This is because considering the fluctuation of the environment enables \texttt{RSR} to be more robust in worse cases.

\textbf{Impact of renting price of SC models}.
Fig. \ref{fig:figure2} shows the profit of customers given different renting prices for the same SC model. Results show that \texttt{RSR} algorithm achieves a maximum higher profit of $16.55\%$ compared to StaticEnv-Opt algorithm and achieves a maximum higher profit of $35.40\%$ compared to Greedy algorithm.

\textbf{Impact of energy budgets}.
We indicate the profit of customers given different energy budgets in Fig. \ref{fig:cust_energy}. Results show that \texttt{RSR} algorithm achieves a maximum higher profit of $19.96\%$ compared to StaticEnv-Opt algorithm and achieves a maximum higher profit of $26.09\%$ compared to  Greedy algorithm. 




\subsection{Seller's Profit Evaluation.}
We test \texttt{IMP} algorithm in various situations. We discuss the profit of the seller with different settings and compare it with the following near-optimal baseline. The allowable range for $\zeta$ is set to $\zeta \in [10, 1000]$. The total number of iterations of \texttt{IMP} algorithm is set to 150.

\begin{itemize}
    \item \textit{Near Optimal (Near-Opt) Algorithm}: Divide the allowable price range evenly into many small intervals and then exhaustively enumerate the values of the boundary points of each interval to calculate the income of the seller, and take the corresponding price when the income is the largest as the final pricing. The default number of intervals is set to 1000.    
\end{itemize}

\textbf{Impact of number of customers}.
Fig. \ref{fig:sub4_1} illustrates the profit of seller when there are different number of customers. More specifically, we discuss cases when the number of customers is 1,  3, and 5. From the results we can find that the seller's profit increases with the increase in the number of customers. Compared to Near-Opt baseline, \texttt{IMP} algorithm has a profit loss of $0.67\%$ with 1 customer, achieves nearly equal profit when with 3 customers, and a profit loss of $1.01\%$ with 5 customers.

\begin{figure}[t]
  \centering
  \begin{minipage}[t]{0.24\textwidth}
    \centering
    \includegraphics[width=\textwidth]{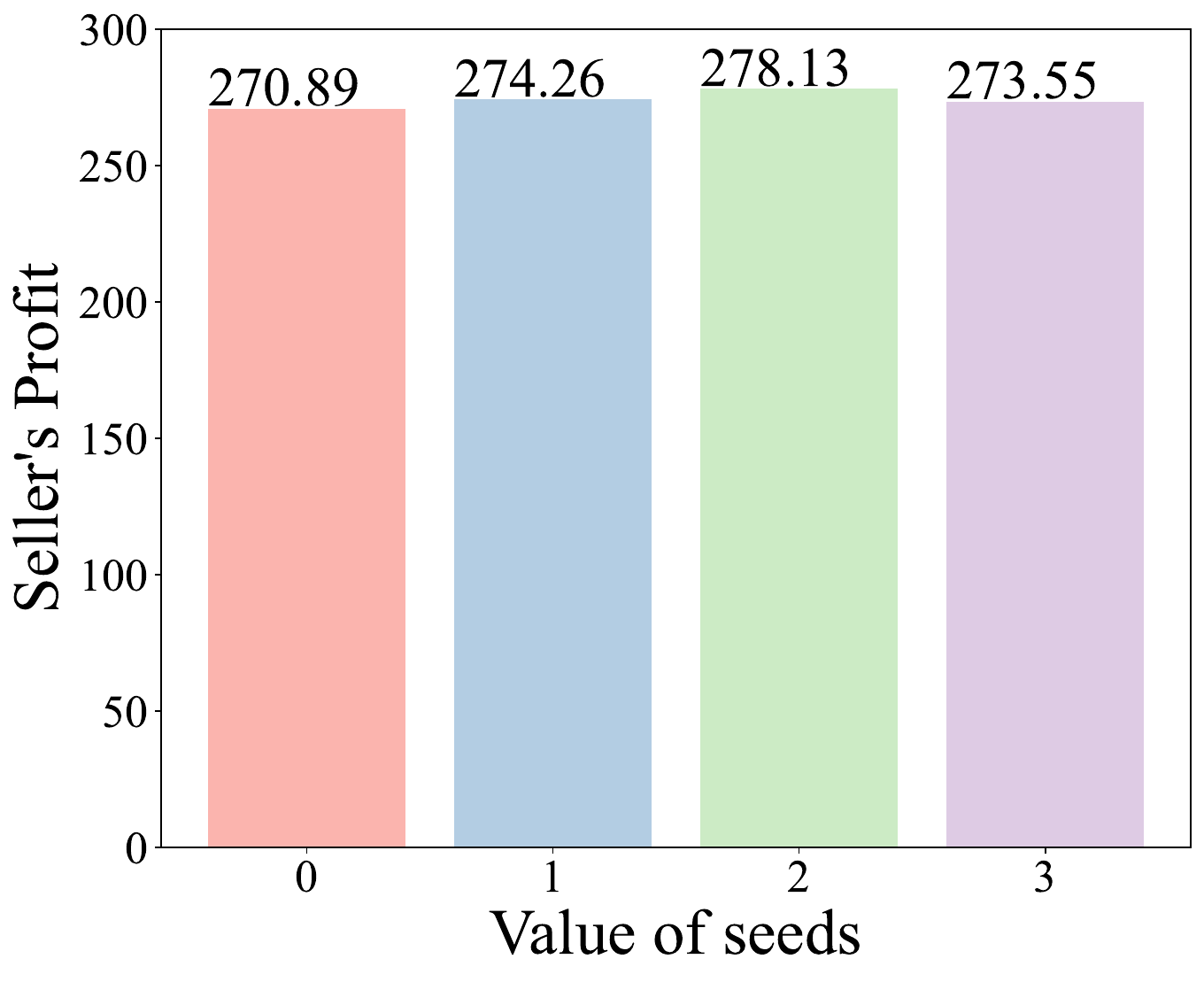}
    \caption{Profit of seller with different random seed}
    \label{fig:seed}
  \end{minipage}
  \hfill
  \begin{minipage}[t]{0.24\textwidth}
    \centering
    \includegraphics[width=\textwidth]{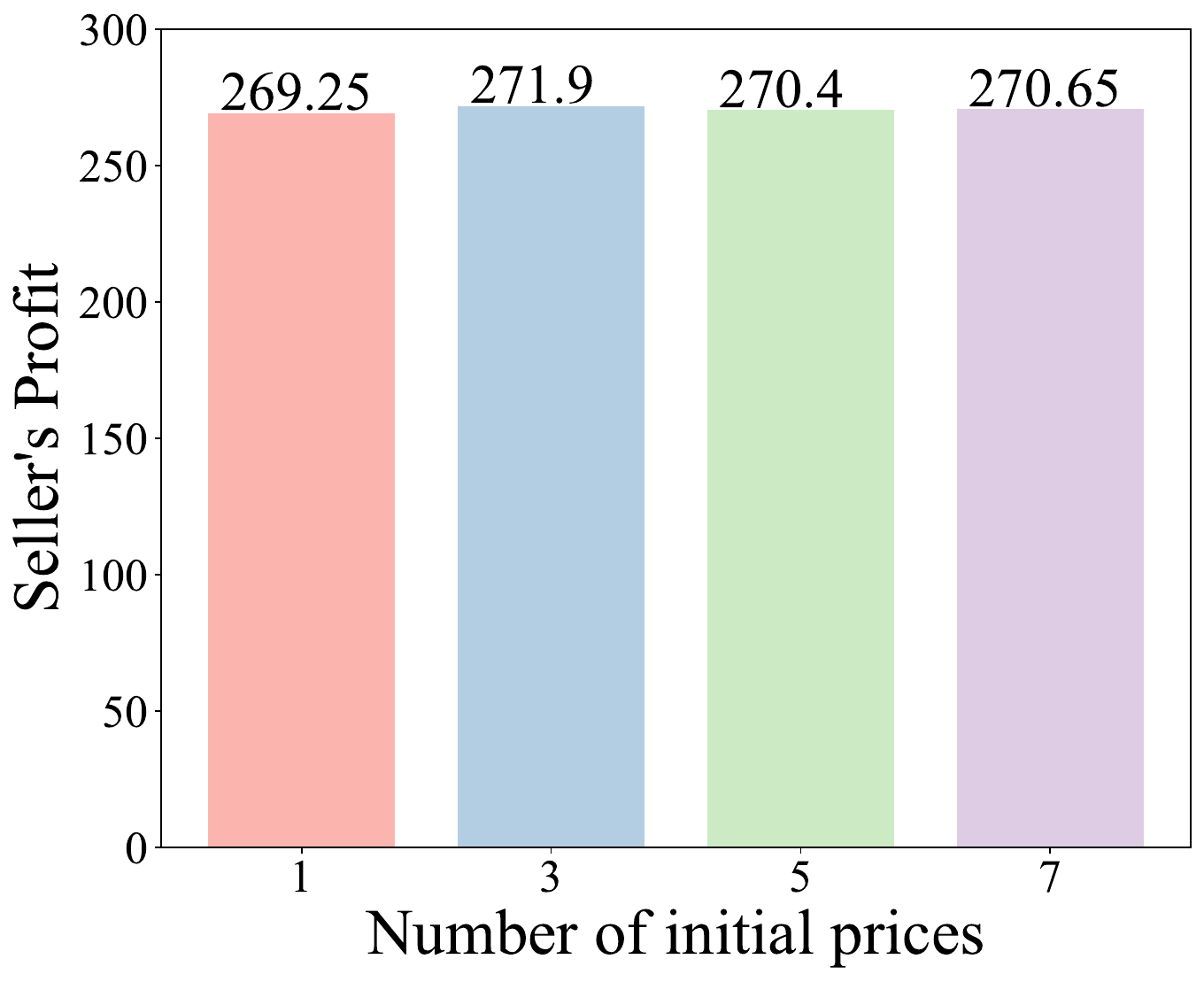}
    \caption{Profit of seller with different \# of initial prices}
    \label{fig:num_start}
  \end{minipage}
  \vspace{-0.3cm}
\end{figure}

\textbf{Impact of the number of available SC models}.
We show the profit of the seller in various number of available SC models in Fig. \ref{fig:sub4_2}. Specifically, we discuss the case when number of SC models in the market are 1, 2, and 3. \texttt{IMP} algorithm has a performance loss of $0.45\%$, $0.25\%$, and $0.47\%$ compared to Near-Opt baseline when available SC model numbers are 1, 2, and 3 respectively.

\textbf{Impact of semantic features transmission distances}.
Fig. \ref{fig:sub4_3} indicates the profit of the seller under different semantic features transmission distances. We vary the average distance between the devices of customers and corresponding base stations to be 750, 1250, and 1750 meters and analyze the resulting profit of the seller. \texttt{IMP} algorithm has a profit loss of $0.09\%$-$9.2\%$ compared to Near-Opt algorithm. Our findings indicate that the profit of the seller decreases as the average distance increases. This is due to the fact that when the semantic features transmission distance is shorter, customers have lower transmission costs and are willing to rent SC models for longer periods of time under the same budget, resulting in increased collected revenue from customers. The SC models selection of customers can result in different costs to the seller in SFT, RM, and RL phases. Longer semantic features transmission distance may lead to customers choosing inferior SC models thereby reducing the training cost of the seller. So, It is not always the case that the profit of the seller increases monotonically with the decrease of semantic features transmission distances of customers.

\textbf{Impact of the value of random seed}.
Fig. \ref{fig:seed} illustrates the profit of the seller under different random seeds. More specifically, we select 4 different seeds and calculate the profit of the seller. It can be observed that the profit has a maximum fluctuation of $1.43\%$ compared to the mean value of all the profits which illustrates the robustness of \texttt{IMP} algorithm under various seeds.

\textbf{Impact of the number of initial prices}.
Fig. \ref{fig:num_start} illustrates the profit of the seller under different number of initial prices which are set to 1, 3, 5, and 7. We set the total number of iterations to be the same to guarantee the fairness of the comparison. Results show that the \texttt{IMP} can achieve good performance in various number of initial prices. We can observe that the seller has the maximum profit when the initial prices number is 3 and has the minimum profit when it is 1. The maximum profit has a profit gain of $0.98\%$ compared to the minimum profit, which indicates the robustness of \texttt{IMP} under different number of initial prices.

\textbf{Comparison of performance and execution time}.
Table \ref{tab:complexity_comparison} compares the profit of the seller and execution time of \texttt{IMP} algorithm and Near-Opt algorithm. 
We use Near-Opt (1000) to represent the Near-Opt algorithm with 1000 intervals and Near-Opt (4000) to represent the Near-Opt algorithm with 4000 intervals. We can find that \texttt{IMP} algorithm can save $76.54\%$ of execution time with only a performance loss of $0.39\%$ compared to Near-Opt (1000).
Also, \texttt{IMP} algorithm exhibits a maximum profit loss of $0.53\%$ but achieves a maximum execution time saving of $93.05\%$ compared to Near-Opt (4000).

\begin{table}[t]
    \centering
    \caption{Comparison of performance and computation overhead.}
    \label{tab:complexity_comparison}
    \vspace{-0.2cm}
    \begin{tabular}{m{2.9cm}<{\centering}  m{2.2cm}<{\centering}  m{2.2cm}<{\centering}}
        \toprule
        \multirow{2}{*}{\textbf{Algorithm}} & \multicolumn{2}{c}{\textbf{Increase Compared to \texttt{IMP}}} \\
        \cline{2-3}
        & \textbf{Profit of seller} & \textbf{Execution time}\\
        \midrule
        Near-Opt (1000) & 100.39\% & 426.22\%  \\
        Near-Opt (4000) & 100.54\% & 1440.45\%  \\
        \bottomrule
    \end{tabular}
\end{table}

\section{Related Work}

\subsection{Semantic Communication System Design}

Farsad et al.\cite{farsad2018deep} propose an LSTM-based encoder-decoder structure that jointly encodes the source and communication channel, leading to a significant performance gain compared to traditional coding methods that treat these components separately.
Xie et al.\cite{xie2021deep} introduce a transformer-based semantic communication system that employs the large language model Bert to better calculate the semantic similarity between the received text and the original text. The authors also utilize transfer learning to adapt the system to dynamic environments.
In \cite{weng2021semantic}, Weng et al. utilize a squeeze-and-excitation network design for the transmission of speech signals and verified its effectiveness in a real multimedia system.
Xie et al.\cite{xie2020lite} design a special distributed lite semantic communication system with limited power consumption to meet the needs of IoT devices.
Xie et al.\cite{xie2021task} focuse on multi-model transmission and developed a semantic communication system based on Resnet-101 and Bi-LSTM to transmit both image and text information. The system is tested in a visual question answering situation.
Yan et al.\cite{yan2022resource} propose the concept of semantic spectral efficiency (S-SE) and apply it to resource allocation in a circular network environment. Simulation results demonstrate the effectiveness of the proposed approach.

\subsection{Semantic Communication Market Design}

Ismail et al.\cite{ismail2022semantic} study the impact of semantic communication in Metaverse and propose an auction mechanism design based on VCG mechanism. Liew et al.\cite{liew2023economics} propose a deep learning based auction mechanism in a hierarchical trading system of semantic communication models to motivate different SC model providers to share semantic engines. They perform a text transmission case using semantic communication and the proposed incentive mechanisms design.
Liew et al.\cite{liew2022economicsicassp} research on the economic issue of energy  of semantic communication systems especially in the IoT environment. IoT devices act as bidders and use a proposed deep learning auction mechanism to bid for energy from Hybrid Access Point (H-AP) to achieve Optimal Profit, Individual Rationality, and Incentive Compatibility.
Luong et al.\cite{luong2022edge} study the computation, storage, communication, and networking perspective of Metaverse as well as the role of semantic communication and design a deep learning based auction algorithm. They test it on a Metaverse semantic communication system and prove the outstanding performance compared to baselines. 

\subsection{Pricing in The Internet of Things}
Ding et al.\cite{ding2022optimal} use two-stage Stackelberg game and three-stage game to study the optimal pricing in IoT networks in three interaction structures: 1) Coordinated interaction structure; 2) Vertically-uncoordinated interaction structure; 3) Horizontally-uncoordinated interaction structure. Li et al.\cite{li2018optimal} study the optimal pricing for peer-to-peer sharing platforms. They define the concept of 'price of information' and show that a simple uniform price scheme can take the place of complicated pricing algorithms. Zhang et al.\cite{zhang2017cooperative} study the optimal data and tethering pricing in Mobile Crowdsourced Internet Access networks.  Jin et al.\cite{jin2019dynamic} model the  requesters’ competitive task pricing in Mobile Crowd Sensing into a Markov game and propose a computation efficient task pricing algorithm for requesters in the system to motivate more workers to work on their tasks. Mao et al.\cite{mao2019pricing} propose MSimple and MGeneral algorithms to deal with one type of buyers and multiple types of buyers respectively in the IoT data market. They further propose an MPractical algorithm with a logarithmic approximation ratio to improve the speed of solving the profit maximization problem. Results show their superiority compared to baselines.

\section{Conclusions}
In this paper, we propose the concept of Large Model as a Service (LMaaS) and conduct a study to analyze the pricing and rental strategies of semantic communication (SC) models, with a specific focus on the utilization of multimodal large language models. We model the interaction between the seller and the customers as a Stackelberg game. Specifically, we consider a scenario where there is uncertainty in the environment and further propose a robust selecting and renting (\texttt{RSR}) algorithm to optimize the profit of customers in the worst case. Additionally, we propose an Iterative Model Pricing (\texttt{IMP}) algorithm to optimize the seller's profits. Finally, we conduct extensive experiments under various settings and demonstrate the effectiveness and robustness of our proposed algorithms compared to baseline methods.








\bibliographystyle{IEEEtran}
\bibliography{reference.bib}

\end{document}